\documentclass[12pt]{iopart}

\usepackage[english]{babel}

\usepackage{graphicx}
\usepackage[colorlinks=true, allcolors=blue]{hyperref}
\usepackage[braket]{qcircuit}
\usepackage{soul}
\usepackage{iopams}
\usepackage{subcaption}

\begin{document}

\title[Impact of Bivariate Gaussian Potentials on Quantum Walks for Spatial Search]{Impact of Bivariate Gaussian Potentials on Quantum Walks for Spatial Search}

\author{Franklin de L. Marquezino $^{1,2, 3}$ and Raqueline A. M. Santos $^1$}

\address{$^1$ Centre for Quantum Computer Science, University of Latvia, Riga, LV-1586, Latvia\\
$^2$ Duque de Caxias Campus, Federal University of Rio de Janeiro, Duque de Caxias, 25240-005, Brazil\\
$^3$ Department of Computer Science and Systems Engineering, Federal University of Rio de Janeiro, Rio de Janeiro, 22290-240, Brazil}
\ead{franklin@cos.ufrj.br, \{franklin.marquezino, rsantos\}@lu.lv}
\vspace{10pt}

\begin{abstract}
Quantum search algorithms are crucial for exploring large solution spaces, but their robustness to environmental perturbations, such as noise or disorder, remains a critical challenge. We examine the impact of biased disorder potentials modeled by a bivariate Gaussian distribution function on the dynamics of quantum walks in spatial search problems. Building on the Ambainis-Kempe-Rivosh (AKR) model for searching on a two-dimensional grid, we incorporate potential fields to investigate how changes in standard deviation and normalization of the bivariate Gaussian function impact the performance of the search algorithm. Our results show that the quantum walk closely mirrors the AKR algorithm when the standard deviation is small but exhibits a rapid decay in success probability as the standard deviation increases.
This behavior demonstrates how the bivariate Gaussian can effectively model a noisy oracle within the AKR algorithm. Additionally, we compare the AKR-based model with an alternative quantum walk model using a Hadamard coin and standard shift. These findings contribute to understanding the robustness of quantum walk search algorithms, and provide insights into how quantum walks can be applied to optimization algorithms.
\end{abstract}

%

%
%
%
%

\section{Introduction}

Quantum walks (QWs), inspired by classical random walks but governed by quantum principles such as superposition and interference, have gained significant attention for their role in quantum search algorithms~\cite{portugal_spatial_2018}.
They offer a flexible framework for modeling and solving spatial search problems, outperforming their classical counterparts. Beyond several kinds of search problems~\cite{zylberman_dirac_2021, fredon_quantum_2022, shenvi_quantum_2003, ambainis_coins_2005}, quantum walks have also been successfully applied to graph traversal~\cite{childs_exponential_2003}, optimization~\cite{varsamis_quantum_2023, liliopoulos_discrete-time_2024}, and even machine learning~\cite{varsamis_quantum_2023a}, demonstrating their versatility across different computational domains. Quantum walks can be classified into two types: continuous-time and discrete-time. In this paper, we focus on discrete-time quantum walks (DTQWs).

Spatial search problems, where the goal is to find a marked vertex in a graph, have been extensively studied in quantum computing. The seminal work by Ambainis, Kempe, and Rivosh (AKR)~\cite{ambainis_coins_2005} introduced a quantum search algorithm for a two-dimensional grid capable of locating a marked vertex efficiently. This algorithm employs the Grover coin and the flip-flop shift operators to govern the evolution of the quantum walker on a grid, achieving significant speedup compared to classical approaches. The AKR algorithm laid the groundwork for a broad class of quantum walk-based search algorithms applicable to different graph topologies and search problems~\cite{portugal_spatial_2018}.


Experimental implementations of continuous-time and discrete-time quantum walks have been achieved across various platforms, including cold atoms~\cite{dadras_experimental_2019,delvecchio_quantum_2020} and photonic systems~\cite{sansoni_two-particle_2012,crespi_anderson_2013, tang_experimental_2018}. These systems offer robust and versatile frameworks for investigating quantum walk dynamics, enabling the study of phenomena such as biased walks, noisy environments, and search algorithms. The advancements in these experimental setups provide not only validation of theoretical predictions but also a practical pathway for implementing the potential-modified quantum walks proposed in this work. 

Quantum walks have been modified to include potential fields, broadening their applicability to problems that involve interaction with external forces. 
For example, Refs.~\cite{genske_electric_2013, cedzich_propagation_2013, bru_electric_2016, arnault_quantum_2020} explored quantum walks in electric fields on one- and two-dimensional lattices. In these studies, the electric field is modeled by a position-dependent phase applied at each step of the walk. Other works have studied spatial search in the context of the Dirac quantum walk with electric fields~\cite{zylberman_dirac_2021, fredon_quantum_2022}, where the quantum walker interacts with the Coulomb field generated by a charge located at the center of the grid.

Building on the work of Genske \etal~\cite{genske_electric_2013}, Varsamis \etal~\cite{varsamis_hitting_2022} numerically studied quantum and random walk hitting times in one- and two-dimensional spaces under specific potential fields. In this model, the walker starts at a chosen vertex and seeks to reach a target vertex with the highest potential.
In a follow-up study, Varsamis \etal\cite{varsamis_quantum_2023} describes
the same quantum walk model where the walker carries a ``charge'' affected by external potentials, with the walk driven by both constant and time-varying potentials. They also observed quantum tunneling through potential barriers and examined the behavior of the walk under various potential distributions.

In addition, quantum walks with potentials have been applied to practical problems. For instance, Varsamis \etal~\cite{varsamis_quantum_2023a} used quantum walks to address peptide and protein folding prediction. Liliopoulos \etal~\cite{liliopoulos_discrete-time_2024},  developed a discrete-time quantum walk-based optimization algorithm where the objective function of the optimization problem is used as an external potential affecting the evolution of the quantum walk. This algorithm consists of two phases: in the first phase, the quantum walk explores a Von Neumann grid starting from a superposition over all vertices, while in the second phase (the intensification phase), the search is refined within a smaller space where the quantum walk is initialized in the center. Their algorithm is applied to train a neural network for binary classification.

Introducing a phase operator to the quantum walk can model decoherence in the system.
Several studies have examined the effects of phase defects in quantum walks. For example, Refs.~\cite{ahlbrecht_asymptotic_2012,lam_ramsauer_2015, zeng_discrete-time_2017, asboth_topological_2020} investigated phase disorder in quantum walks on one- and two-dimensional grids, while Refs.~\cite{zhang_one-dimensional_2014, zhang_two-dimensional_2014, farooq_quantum_2019} explored the impact of phase noise inserted into the shift operator. 
Furthermore, Abal \etal~\cite{abal_decoherence_2009} studied the introduction of phase noise into the coin operator for both the hypercube and two-dimensional grid.

In this paper, we explore how biased disordered potentials, specifically a bivariate Gaussian distribution function acting as a potential field, influence the behavior of quantum walks in spatial search problems. 
Our study builds on the AKR-based model and introduces potential fields to observe how varying the standard deviation and normalization of the bivariate function impacts the algorithm's performance.
Our numerical results demonstrate that when the standard deviation is small, the quantum walk exhibits success probabilities similar to the AKR algorithm. However, as the standard deviation increases, we observe a specific range in which the success probability decays rapidly, with this behavior consistent across different grid sizes. This highlights the sensitivity of the quantum walk to changes in the potential's parameters and demonstrates that the bivariate Gaussian function can effectively model a noisy oracle within the AKR algorithm.
We also compare the AKR-based model to an alternative quantum walk model (using a Hadamard coin and standard shift) to highlight the robustness of quantum walk-based search algorithms under different conditions.

This paper is structured as follows. In Sec.~\ref{sec:electric-qw}, we review the spatial search algorithm and the model of quantum walks with electric fields. In Sec.~\ref{sec:search}, we describe the quantum walk model on the two-dimensional grid which uses the bivariate Gaussian distribution as a potential function. In Sec.~\ref{sec:numerical}, we present our numerical simulations and results. Finally, in Sec.~\ref{sec:conclusion}, we present our conclusions and final discussions.

\section{Spatial search and potential fields}
\label{sec:electric-qw}

In the spatial search problem, we have a graph $G=(V,E)$, and we want to find a vertex $v \in V$ that satisfies a certain property. We assume that there is an oracle capable of testing whether any vertex of the graph satisfies the property. However, after each test, we must move to a neighbor vertex before using the oracle again. The graph $G$ can be used to represent several situations in which the relationship between elements of the search space are relevant, such as their disposition in memory. For example, if $G$ is the complete graph with $N$ vertices, then this means that all elements are immediately accessible in memory after each step, so the spatial aspect of the search is not relevant. In this case, a classical algorithm needs $O(N)$ oracle queries, while the well-known Grover's quantum algorithm needs $O(\sqrt{N})$ oracle queries~\cite{grover_fast_1996}. If $G$ is the $n$-dimensional hypercube, then we have a quantum algorithm that requires $O(\sqrt{2^n})$ oracle queries, versus the best classical algorithm requiring $O(2^n)$ queries, which also represents an optimal quantum speedup~\cite{shenvi_quantum_2003}.

There is a general framework for describing quantum walk-based search algorithms. Let $\ket{\psi_0}$ be the starting state and $M$ be the set of marked states. Then we have to define two unitary transformations $R$ and $U$. The first transformation 
\begin{equation}
 R = I-2\sum_{m \in M}\ket{m}\bra{m} 
\end{equation}
acts as an oracle inverting the phases of basis states corresponding to searched vertices. The second transformation $U$ is the quantum walk evolution associated with the graph. In the discrete-time model of quantum walk, the evolution operator is written as $U = S(C \otimes I)$, where $S$ is the shift operator and $C$ is the coin operator.
We have a lot of flexibility to choose those operators. The coin can be any unitary operator, and the shift operator is restricted only by the topology of the graph. In general, $S\ket{e,v} = \ket{e^\prime, v^\prime}$, where $v^\prime$ is the vertex connected to $v$ by edge $e$. 

The quantum search algorithm consists of applying $(UR)^t\ket{\psi_0}$ and then measuring the walker's position. The number of steps $t$ depends on the particular instance of the search problem.
Under a few assumptions, it is possible to prove the correctness of this search algorithm and also analyze its complexity. For a complete review of the search algorithms based on quantum walk, including the complete analysis of correctness and complexity, refer to Portugal~\cite{portugal_spatial_2018}.

For the present work, the quantum search algorithm on the two-dimensional grid will be particularly useful. The standard quantum algorithm for spatial search on a two-dimensional grid is due to Ambainis, Kempe and Rivosh (AKR)~\cite{ambainis_coins_2005}. In the AKR algorithm---and in most quantum walk-based search algorithms---the coin operator $C$ should be the so-called Grover operator, i.e., the matrix given by $C = 2\ket{d}\bra{d} - I,$ where
$ \ket{d} = H^{\otimes 2}\ket{0}$ is the uniform superposition of coin states. The shift operator should be the so-called flip-flop shift, defined as
\begin{equation}
S\ket{j,k}\ket{x, y} = \ket{j,1-k}\ket{x + j(-1)^k , y + (1-j)(-1)^k},
\end{equation}
where the operations inside the position ket are performed modulus $N$ to implement periodic boundary conditions.

Genske \textit{et al.} described a model of a discrete-time quantum walk on the one-dimensional lattice subject to an artificial electric field~\cite{genske_electric_2013}. This model starts with the standard quantum walk, which evolves according to the above-mentioned operator $U = S ( C \otimes I)$. The walk is then turned into an electric one by adding an extra operation $\hat{F}_E = \exp(i \hat{x} \Phi)$, where $\hat{x}$ is the lattice position operator, so that $\exp(i \hat{x} \Phi)\ket{c, x} = \exp(i x \Phi)\ket{c, x}$. The constant $\Phi$ is called the electric field. $\hat{F}_E$ performs a phase shift depending linearly on $x$, which is imprinted by an electric field between two adjacent lattice sites. The overall evolution is then given by $U_\Phi = \hat{F}_EU$. In this way, the dynamics of the quantum walk behaves as if there were an electric field acting on a charged particle in a lattice. Later works have also analyzed quantum walks with non-linear potential fields, which can have important applications to optimization problems~\cite{varsamis_quantum_2023}.

\section{Quantum search algorithms in potential fields}
\label{sec:search}

As we discussed in the previous section, the approach to solving spatial search problems with quantum walk-based algorithms is very general and can be applied to a very broad class of graphs~\cite{ambainis_quadratic_2020, apers_quadratic_2022}. However, for simplicity, we consider the two-dimensional grid in the present work.
Notice that the oracle operator from the original AKR algorithm behaves roughly as a delta potential well centered at the searched vertex---more precisely, instead of being infinite, the potential values $\pi$ on the searched vertex and zero elsewhere. 

The bivariate Gaussian distribution is the generalization of the Gaussian distribution to two dimensions, and the general form of its probability density function is given by
\begin{equation}
f_{\sigma, \mu}(x,y) = \frac{1}{2\pi \sigma_X \sigma_Y \sqrt{1-\rho^2}}\exp\left\{-\frac{z}{2(1-\rho^2)}\right\},
\end{equation}
where
\[
z \equiv \frac{(x-\mu_X)^2}{\sigma_X^2} + \frac{(y-\mu_Y)^2}{\sigma_Y^2} - 2\rho\frac{ (x-\mu_X)(y-\mu_Y)}{\sigma_X \sigma_Y},
\] 
and parameters $\mu_X, \mu_Y$ are the means, $\sigma_X, \sigma_Y$ are the stardard deviations, and $\rho$ is the correlation of random variables $X$ and $Y$.

\begin{figure}
    \centering
    \includegraphics[width=1.0\textwidth]{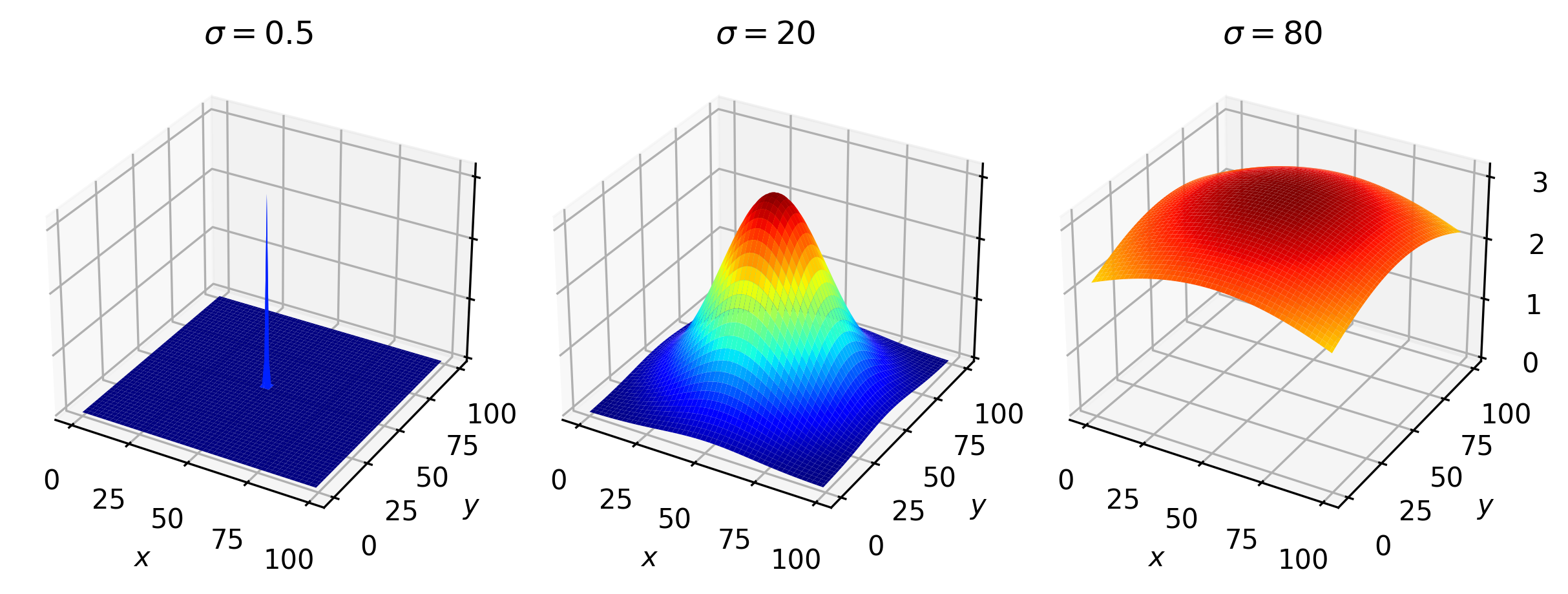}
    \caption{Surface plots of the bivariate Gaussian distribution for different standard deviations, normalized so that the maximum value is equal to $\lambda = \pi$.  (Color online.)}
    \label{fig:bivariate}
\end{figure}

In Fig.~\ref{fig:bivariate}, we have examples of the bivariate Gaussian distribution for different standard deviations. In all three plots, we use the center of the $\sqrt{N}\times \sqrt{N}$ two-dimensional lattice as the mean, and we consider that both $X$ and $Y$ variables are not correlated. We also normalize the function so that the maximum value is equal to $\lambda$, which is very convenient for our comparison with search algorithms. Therefore, we are interested in distribution 
\begin{equation}
    \hat{f}_{\sigma, \mu}(x,y) =  \frac{\lambda f_{\sigma, \mu}(x,y)}{\max_{x,y}f_{\sigma, \mu}(x,y)},
\end{equation}
with $\sigma = \sigma_X = \sigma_Y$, and $\rho=0$, and $\mu = \mu_X = \mu_Y = \sqrt{N}/2$.

Notice that as the standard deviation approaches zero, the distribution becomes the usual oracle used in standard quantum walk-based search algorithms, in which the central vertex of the lattice is the marked element. As the standard deviation increases, the distribution becomes more spread out and flattened to the point that the phase change becomes approximately the same to all vertices, which means that none of them are actually marked.

In this work, we consider a quantum walk in a two-dimensional grid with flip-flop shift operator and bivariate Gaussian potential field. Therefore, the evolution operator considered in the following sections is given by
\begin{equation}
    U\ket{j,k}\ket{x,y} = e^{i \hat{f}_{\sigma, \mu}(x,y)}  S (C \otimes I) \ket{j,k}\ket{x,y}. 
\end{equation}
If we denote a generic state of the walker at time $t$ by the unit vector
\begin{equation}
\ket{\Psi(t)} = \sum_{j,k=0}^{1}\sum_{x,y=0}^{\sqrt{N}-1} \psi_{j,k;x,y}(t) \ket{j,k}\ket{x,y},
\end{equation}
then the probability of finding the walker at position $(x,y)$ if measured at that instant can be given by the Born rule,
\begin{equation}
    p_t(x,y) = \sum_{j,k=0}^{1} \| \psi_{j,k;x,y}(t) \|^2.
\end{equation}

\section{Numerical simulations}
\label{sec:numerical}

In this section, we examine the behavior of a quantum walk driven by a bivariate Gaussian function acting as a potential. Our aim is to assess the algorithm's ability to search for a vertex with the highest potential, specifically, the central vertex in the grid. The bivariate function behaves like an oracle when its standard deviation $\sigma$ is close to $0$. Here, we analyze how modifying $\sigma$ and the normalization of the potential function affects the performance of the quantum walk. Additionally, we compare two distinct models of quantum walks to highlight their respective advantages.

\subsection{Varying the standard deviation $\sigma$ of the bivariate function}

\begin{figure}
    \centering
    \includegraphics[width=0.9\textwidth]{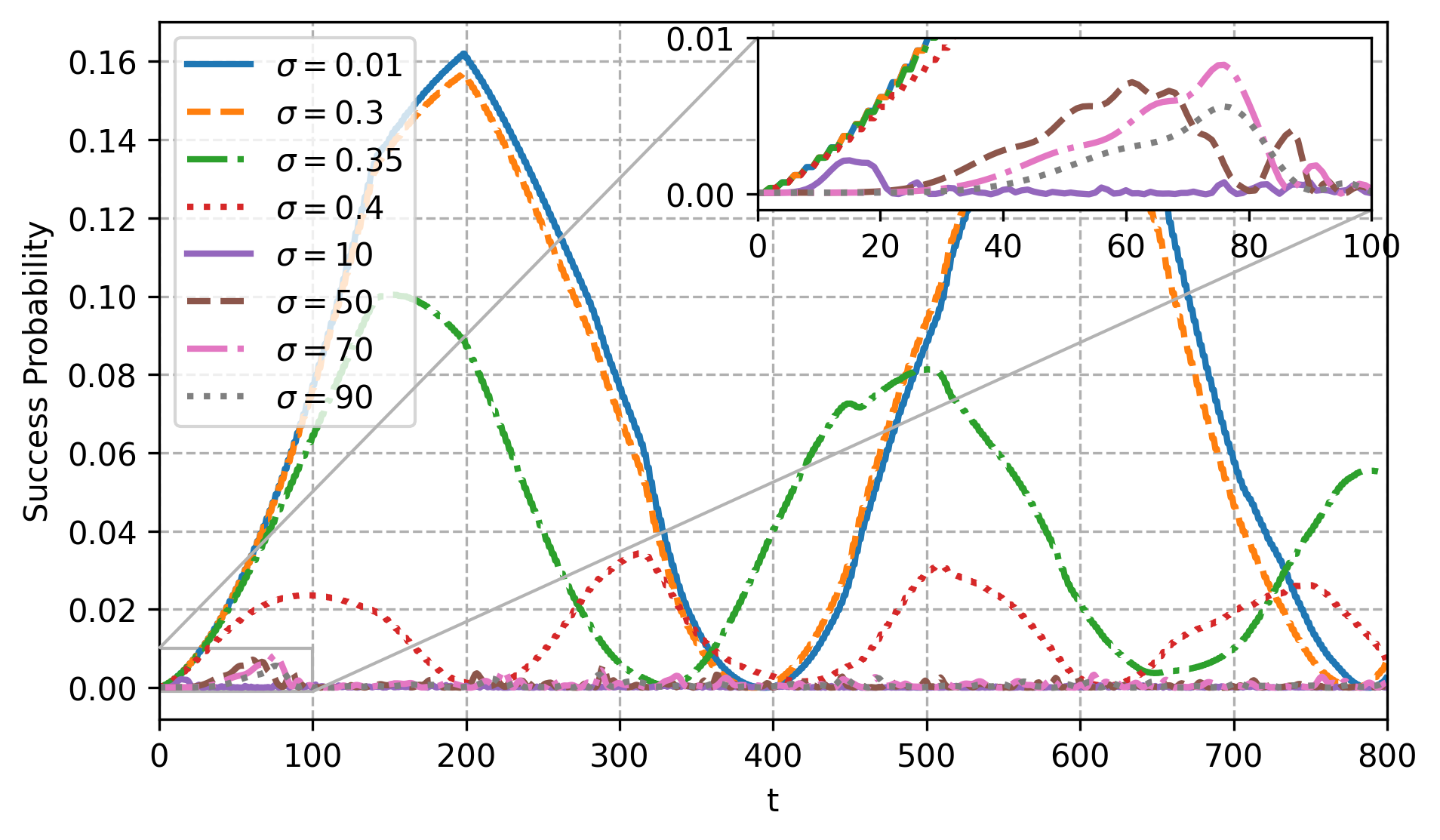}
    \caption{Success probability over time for a $100\times 100$ grid for various values of $\sigma$. At $\sigma = 0.01$, the curve mirrors the original oracle-based quantum search. As $\sigma$ increases, the success probability decreases and eventually converges to the initial probability. (Color online.)}
    \label{fig:success-prob}
\end{figure}

We begin by fixing the normalization value at $\lambda = \pi$, and apply the standard AKR quantum walk configuration: Grover coin, flip-flop shift, and periodic boundary conditions.
The success probability of the quantum walk is significantly influenced by varying values of $\sigma =\sigma_x=\sigma_y$ in the bivariate Gaussian distribution, as illustrated in  Fig.~\ref{fig:success-prob}. We plot the success probability over time for a $100\times 100$ grid and different $\sigma$ values. The success probability corresponds to the probability at the vertex $(50,50)$, which has the highest potential value. 

When $\sigma=0.01$,  the quantum walk behaves similarly to the original AKR algorithm, with a high success probability at the central vertex. As $\sigma$ increases, the success probability begins to decline. For the values of $\sigma$ shown in Fig.~\ref{fig:success-prob}, the success probability still reaches higher values than the uniform distribution but gradually converges toward the initial uniform value as $\sigma$
increases further. This behavior indicates that the quantum walk becomes progressively less effective at identifying the vertex with the highest potential as the potential distribution spreads more broadly across the grid.

\begin{figure}
    \centering
    \includegraphics[width=\textwidth]{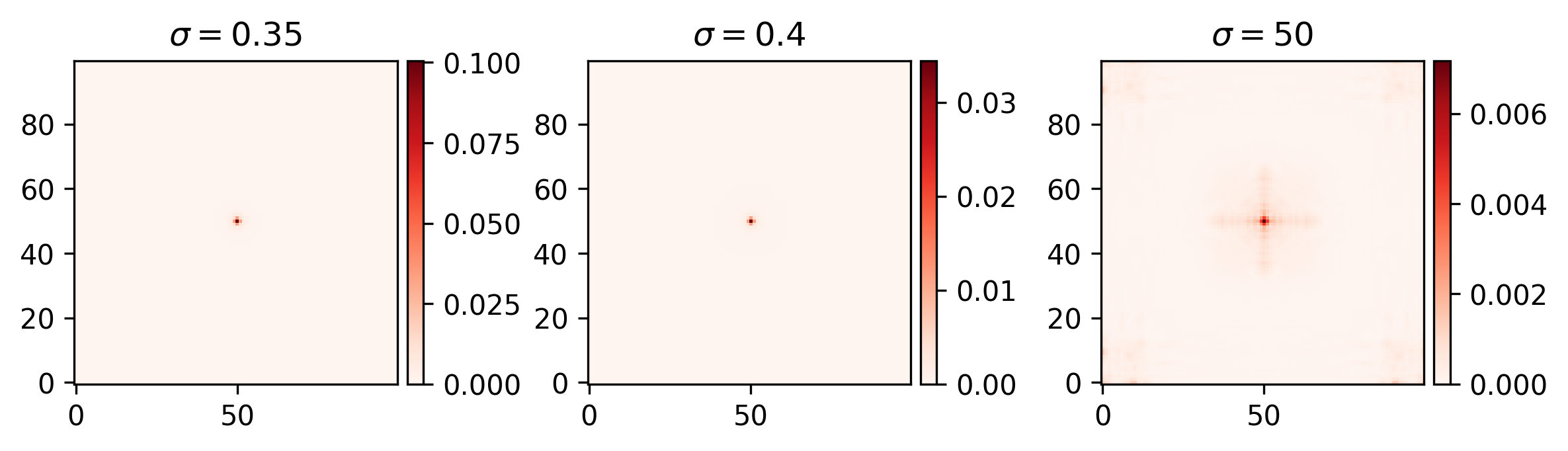}
    \caption{Probability distributions over a $100\times 100$ grid at $t=153$ for $\sigma = 0.35$; $t=315$ for $\sigma=0.4$; and $t=61$ for $\sigma = 50$. The highest probabilities occur near the center of the grid, aligning with the behavior of the bivariate function. (Color online.)}
    \label{fig:pds}
\end{figure}

Next, in Fig.~\ref{fig:pds}, we illustrate the probability distributions at time step $t=153, 315, 61$, corresponding to the maximum success probability achieved by the quantum walk with $\sigma = 0.35, 0.4, 50$. The maximum success probability, $p_{max}$, is calculated in the interval $[0,3\sqrt{N}]$. As expected, the highest probabilities are concentrated near the center of the grid, consistent with the peak of the bivariate Gaussian distribution. The maximum probability at the target vertex is approximately 0.1, 0.034 and 0.00716, respectively.

\begin{figure}
    \centering
    \includegraphics[width=0.9\textwidth]{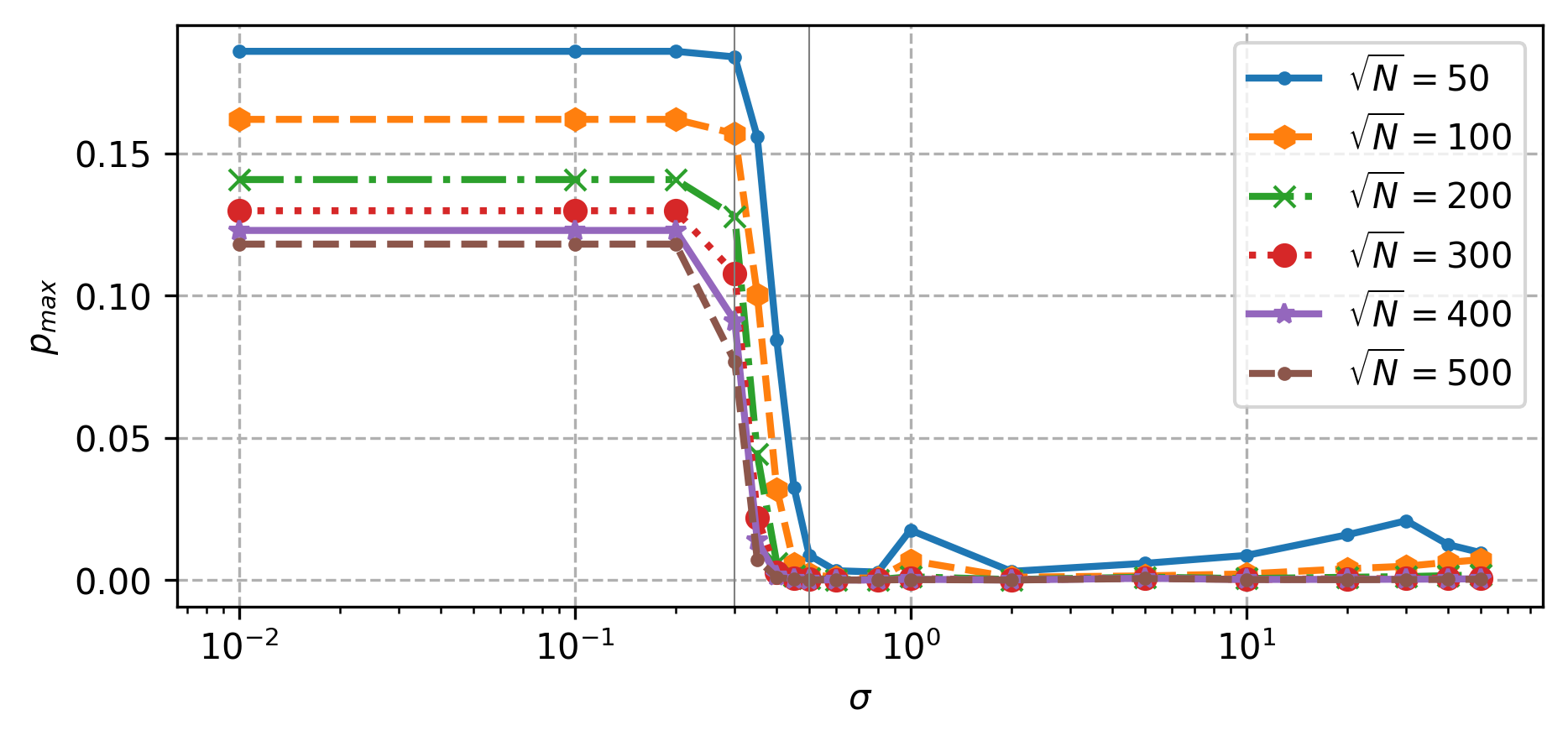}
    \caption{Maximum success probability for different values of $\sigma$ (in log scale) across grid sizes. A step drop in probability is observed around $\sigma \in [0.3,0.5]$. For larger $\sigma$, the probability approaches the initial uniform probability. (Color online.)}
    \label{fig:max-sigmas}
\end{figure}

To further explore the role of $\sigma$, we obtain the maximum success probability for various values of $\sigma$ and grid sizes, as shown in Fig.~\ref{fig:max-sigmas}. Interestingly, there is a sharp decrease in success probability within the interval $\sigma \in [0.3,0.5]$, suggesting that the quantum walk performs best when $\sigma$ is small and the function approaches a delta function. For large $\sigma$, the algorithm’s behavior increasingly resembles sampling from a uniform distribution, which can be useful for understanding the effects of noise in an oracle-based quantum search.

\begin{figure}
\centering
    \begin{subfigure}[t]{\textwidth}
         \centering
         \includegraphics[width=0.9\textwidth]{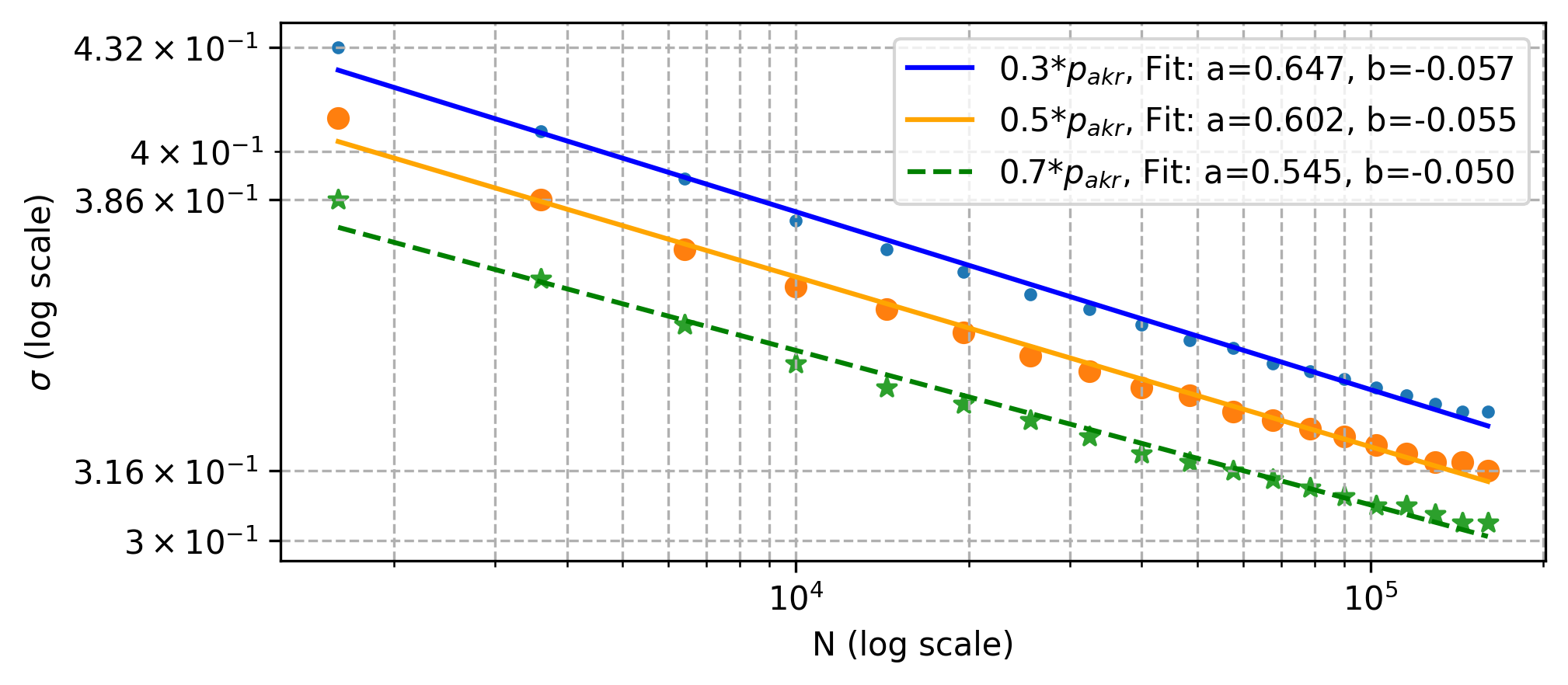}
         \caption{}
         \label{fig:sigma-first}
     \end{subfigure}\\
    \begin{subfigure}[t]{\textwidth}
         \centering
         \includegraphics[width=0.9\textwidth]{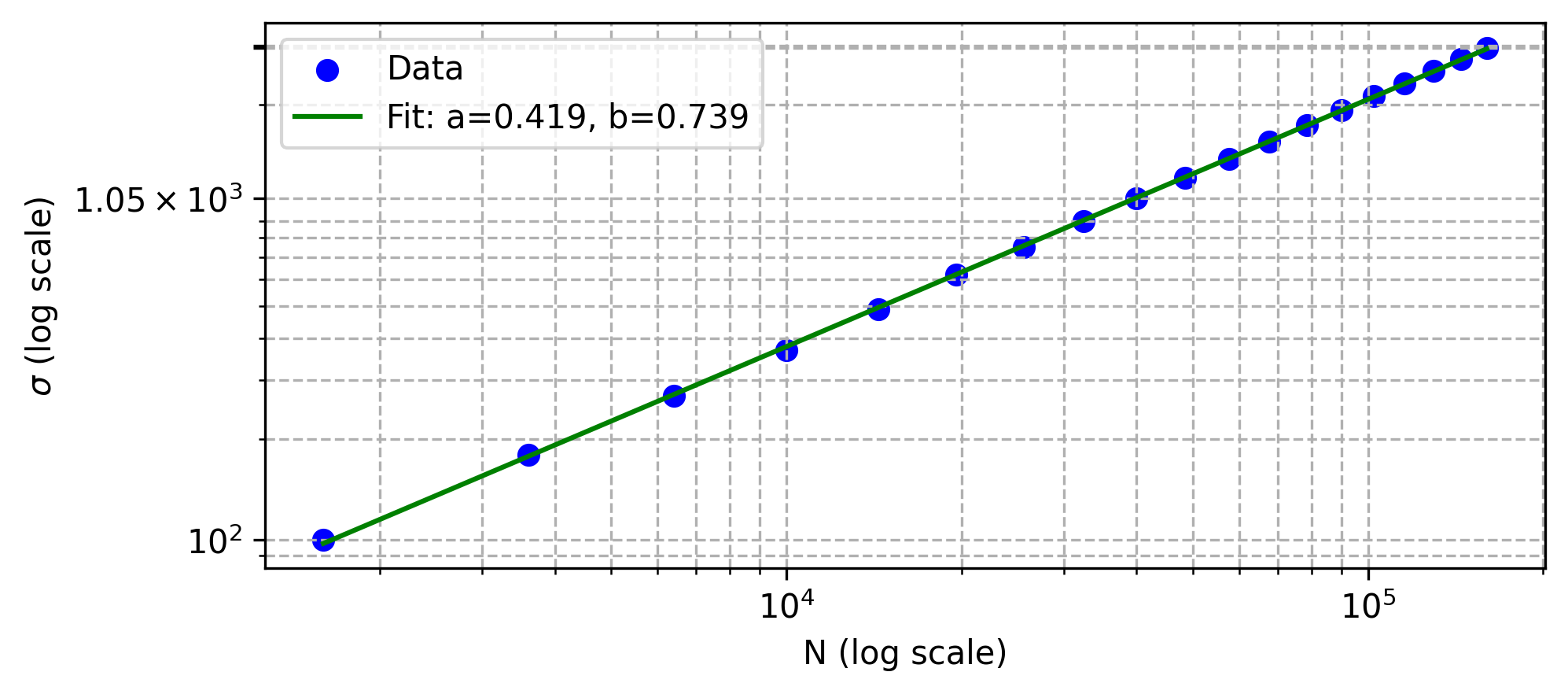}
         \caption{}
         \label{fig:sigma-second}
     \end{subfigure}
    \caption{(a) Minimum $\sigma$ for which the maximum success probability gets below a certain threshold as a function of the grid size. (b) Minimum $\sigma$ for which the maximum success probability gets $\epsilon = 0.5$ close to the uniform distribution. (Color online.)}
    \label{fig:sigmas-regimes}
\end{figure}

In our numerical simulations, we observed that the variance $\sigma$ in which the probability of success suddenly drops has a weak dependence on the size of the grid. Fig.~\ref{fig:sigma-first} captures the smallest variance $\sigma$ for which the maximum success probability  $p_{max}$ of the disordered walk stays below a certain percentage of the maximum success probability $p_{akr}$ of the ideal AKR algorithm, as a function of the grid size. Specifically, it captures the smallest $\sigma$ such that $p_{max}(\sigma) \leq \epsilon p_{akr}$. Interestingly, the step drop in success probability occurs when $\sigma \sim N^{-0.05}$, indicating a weak grid-size dependence. 

After this sharp drop,the quantum walk transitions into an intermediate regime where the success probability remains significantly lower than that of the ideal AKR algorithm but does not yet match the uniform distribution.Fig.~\ref{fig:sigma-second} illustrates the smallest variance $\sigma$ for which the maximum success probability $p_{max}$ of the disordered walk gets $\epsilon$ close to the uniform superposition probability $p_u$, i.e., the smallest $\sigma$ such that $$1-\frac{p_u}{p_{max}(\sigma)} \leq \epsilon.$$ 
We observed that this intermediate regime persists until 
$\sigma \sim N^{0.74}$, beyond which the behavior of the quantum walk fully aligns with uniform sampling. This suggests that the interplay between the potential’s spread and the grid size governs the transition between the coherent search, intermediate, and uniform sampling regimes.

The sharp drop and subsequent regimes can be explained by the nature of the bivariate Gaussian potential. When $\sigma$ is small, the potential closely resembles a delta function, creating a well-defined target that facilitates constructive interference at the marked vertex. As $\sigma$ increases, the potential broadens, reducing its effectiveness in guiding the quantum walk. The intermediate regime reflects a state where the potential is sufficiently spread to weaken interference effects but retains some spatial bias. Finally, as $\sigma$ becomes large, the potential flattens across the grid, causing the quantum walk to lose its directional advantage and behave like uniform sampling. 

\subsection{Varying the normalization $\lambda$ of the bivariate function}

\begin{figure}
\centering
    \begin{subfigure}[t]{0.9\textwidth}
         \centering
         \includegraphics[width=0.9\textwidth]{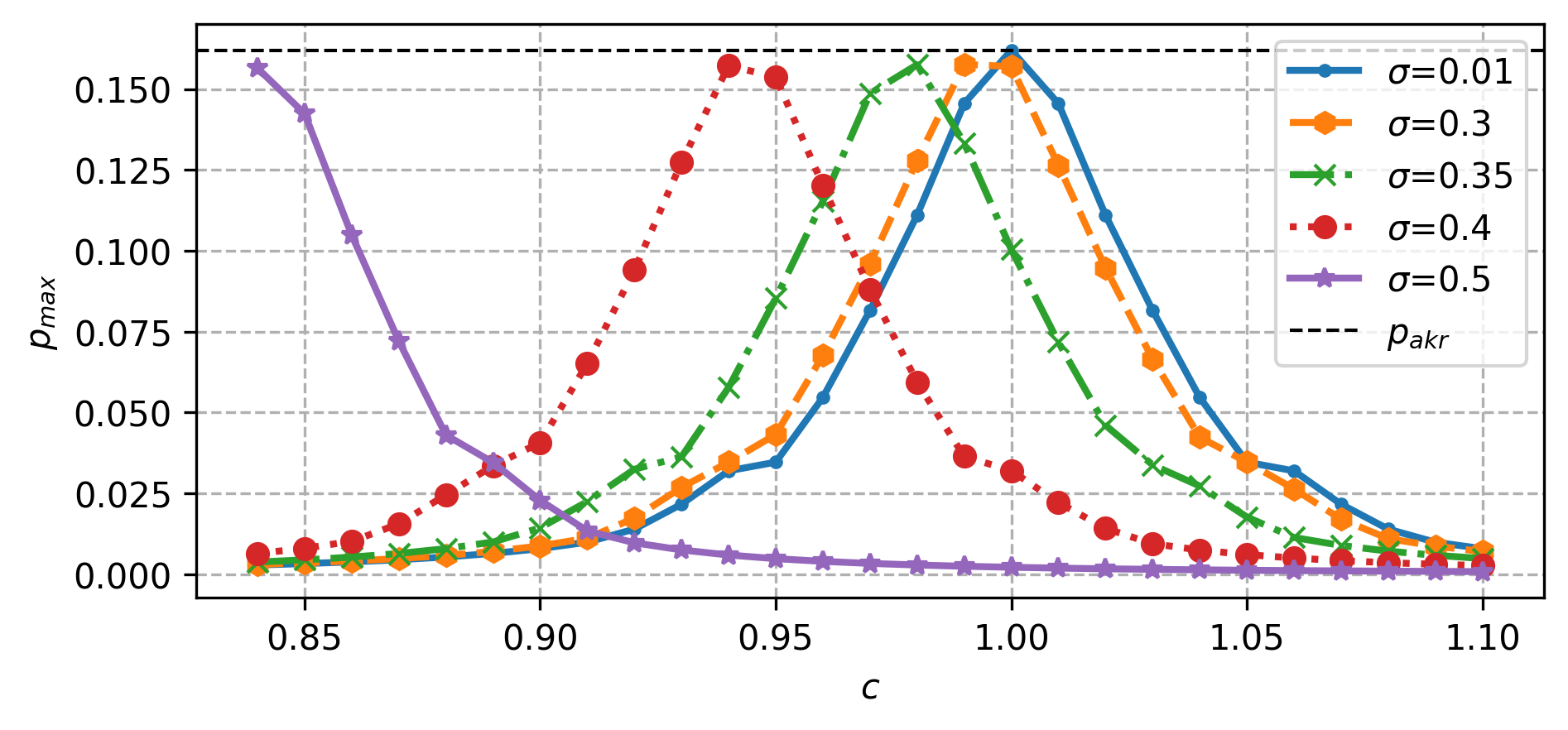}
         \caption{}
         \label{fig:c-closer}
     \end{subfigure}\hfill
    \begin{subfigure}[t]{0.9\textwidth}
         \centering
         \includegraphics[width=0.9\textwidth]{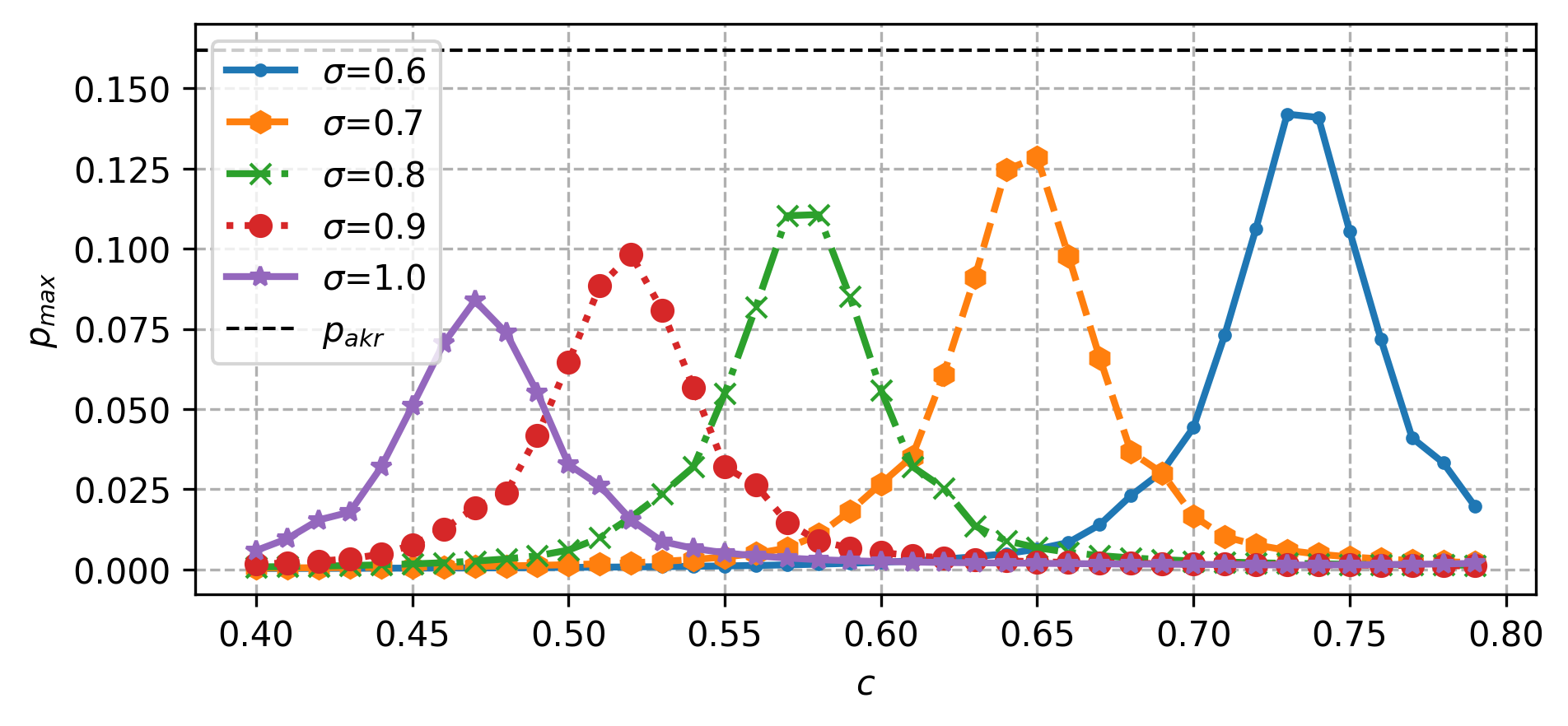}
         \caption{}
         \label{fig:c-not-closer}
     \end{subfigure}
\caption{Maximum success probability for various normalization values $\lambda = c\pi$ and different $\sigma$. The grid size is $100\times 100$. Each curve represents a different value of $\sigma$. The peak probability diminishes and shifts leftward as $\sigma$ increases. (a) For small $\sigma$, the performance can be close to the AKR algorithm (dashed black line). (b) For larger $\sigma$, a more pronounced decrease of the peak probability is observed. The red dotted line is the initial probability. (Color online.)}
\label{fig:vary-c}
\end{figure}

Next, we investigate the impact of adjusting the normalization factor of the bivariate function, expressed as $\lambda = c\pi$. Here, the size of the grid remains fixed at $100\times100$. In Fig.~\ref{fig:vary-c}, we present the maximum success probability for different values of $c$ and $\sigma$. As $\sigma$ increases, we observe that the peak probability shifts leftward and decreases in magnitude. For small values of $\sigma$, as shown in Fig.~\ref{fig:c-closer}, the peak value is still very close to the original value in the AKR algorithm, depicted by the dashed black line. The reduction in the peak becomes more pronounced in Fig.~\ref{fig:c-not-closer}, reinforcing the earlier observation that the search becomes less efficient for larger $\sigma$. Notably, due to the periodic nature of the potential operator, $\exp(i\hat{f})$, it is still possible to fine-tune $\sigma$ and $\lambda$ to achieve performance comparable to the original AKR algorithm.

\subsection{Comparison of quantum walk models}
Finally, we compare the AKR-based model (Model 1: Grover coin, flip-flop shift, periodic boundary conditions) to an alternative quantum walk model (Model 2: Hadamard coin, standard shift, reflective boundary conditions), similar to the optimizer algorithm described in~\cite{liliopoulos_discrete-time_2024}. In this experiment, we apply the bivariate function as a potential operator and compute the maximum success probability for both models over the time interval $[0,300]$ for various values of $\sigma$ in a $100\times 100$ grid.

\begin{figure}
    \centering
    \includegraphics[width=0.9\textwidth]{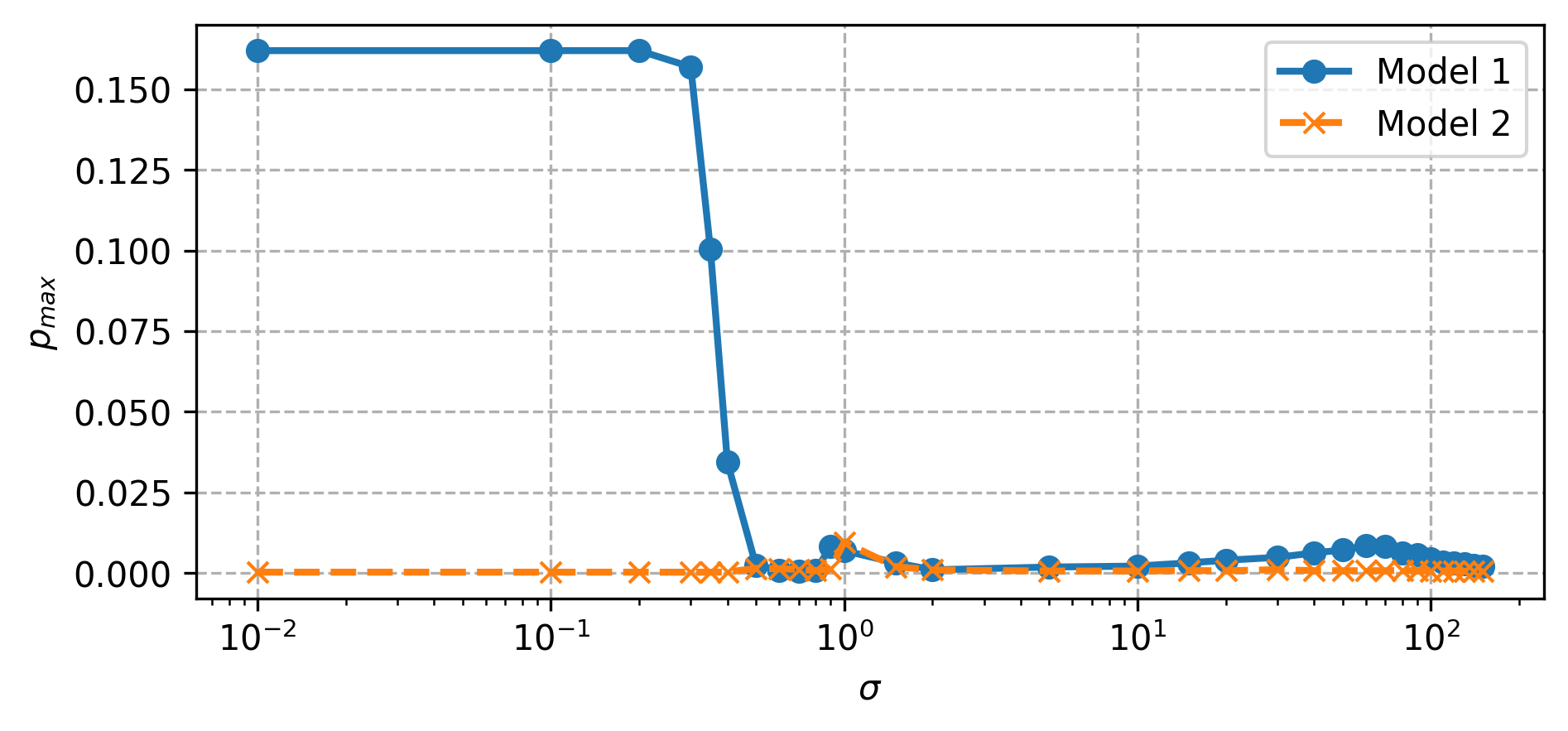}
    \caption{Comparison of success probability for Model 1 (AKR-based model) and Model 2 (Hadamard coin, standard shift, reflective boundaries) across $\sigma$ (in log scale). Model 1 (solid blue curve) consistently outperforms Model 2 (dashed orange curve), which remains near the initial probability. The grid size is $100\times 100$. (Color online.)}
    \label{fig:comparison}
\end{figure}

As seen in Fig.~\ref{fig:comparison}, Model 1 consistently outperforms Model 2, which remains close to the initial probability. This suggests that the AKR-based model is better suited for tasks that involve a well-defined potential function.

\begin{figure}[!htb]
    \centering
    \includegraphics[width=0.9\textwidth]{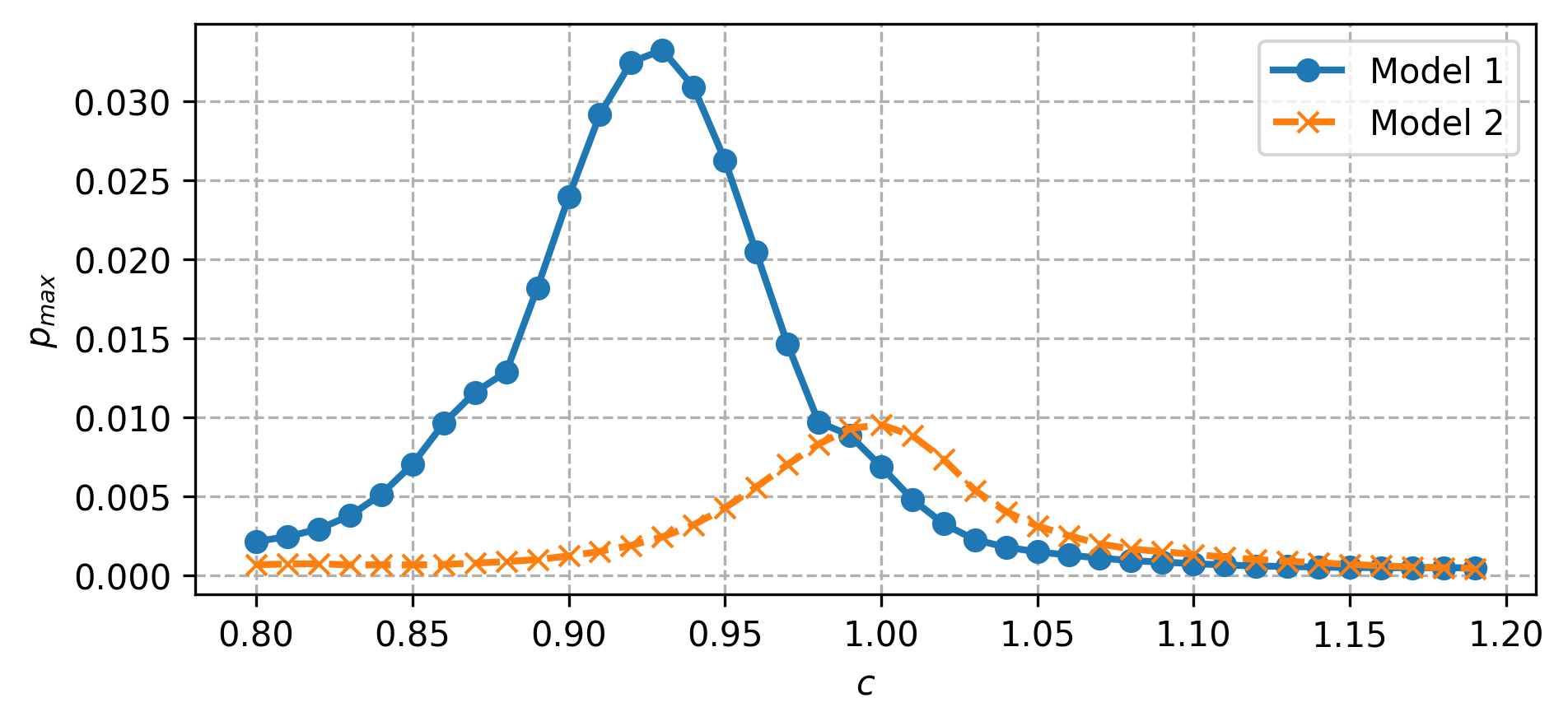}
    \caption{Maximum of the success probability for varying normalization values $\lambda=c\pi$ for a $100\times 100$ grid at fixed $\sigma = 1$. Model 1 (solid blue curve) is the AKR-based model. Model 2 (dashed orange curve) is the configuration with Hadamard coin, standard shift and reflexive boundary conditions. Although Model 2 slightly outperforms Model 1 for some $c$, Model 1 generally provides better performance in different scenarios. (Color online.)}
    \label{fig:comparison2}
\end{figure}

In Fig.~\ref{fig:comparison2} we further explore the effect of different normalization values $\lambda = c\pi$ for $\sigma = 1$. Interestingly, for some values of $c$, Model 2 slightly exceeds Model 1, but overall, Model 1 remains superior in terms of search performance. This suggests that the DTQW optimizer algorithm described by~\cite{liliopoulos_discrete-time_2024}, could benefit from incorporating the AKR-based model for higher success rates.

\section{Conclusion and further discussions}
\label{sec:conclusion}

In this paper, we have explored the performance of quantum walks using a bivariate function as a potential, comparing different configurations and models. When $\sigma$ is close to 0, the function behaves similarly to an oracle, directing the quantum walk toward the vertex with the highest potential. In this regime, the quantum walk behavior closely mirrors that of traditional quantum search algorithms. As
$\sigma$ increases, the sharpness of the potential decreases and the search becomes less focused, transitioning towards behavior resembling random sampling across the grid.

A key insight is that the bivariate Gaussian potential introduces a flexible framework for modeling decoherence and noisy oracles in the AKR algorithm. As the function broadens with increasing $\sigma$, the quantum walk behaves more like uniform random sampling. 
Our numerical approach establishes an initial understanding of the interaction between quantum walks and bivariate Gaussian potentials. Analytical methods will be a crucial next step to derive deeper insights and potentially generalize the observed phenomena.

Our findings indicate that Model 1 (AKR configuration) consistently outperforms Model 2 (Hadamard coin, standard shift) in the context of potential-based quantum searches.
In our simulations, Model 1 seems to show superior performance not only with the bivariate Gaussian function, but also with other potential functions, such as the Ackley and Rastringin functions. These functions feature a global minimum surrounded by numerous local minima, making the search process more challenging. In contrast, Model 2, while less effective in these scenarios, may still offer advantages in specific phases of hybrid algorithms. For example, in the second phase of the optimizer algorithm described in \cite{liliopoulos_discrete-time_2024} — which is an intensification phase — starting from a localized state may favor the Hadamard coin over the Grover coin. This suggests a complementary role for Model 2 in certain problem setups.

Our results open new avenues for exploration. An important future direction is to derive analytical results that can further explain the trends observed in our simulations, particularly the relationship between the normalization factor
$\lambda$, the standard deviation $\sigma$, and the resulting success probabilities. 
While our results focus on a simplified bivariate Gaussian potential, future studies could explore more generalized cases, such as allowing unequal standard deviations or introducing correlations.
Moreover, a systematic investigation into the effects of potential functions beyond the bivariate Gaussian could provide valuable insights into the behavior of quantum walk based algorithms for solving optimization problems.

 While our study focuses on the implications of finding a vertex with the highest potential in our setup, it also opens pathways for exploring the transport and dynamical properties of quantum walks under static disorder. Notably, the behavior of the quantum walk could be analyzed in the context of localization phenomena, where increasing disorder suppresses coherent transport.

The bivariate Gaussian potential introduced here could be realized experimentally by tailoring position-dependent phase shifts in optical systems or manipulating local potentials in cold-atom setups, for instance. These platforms provide a promising basis for extending the applicability of quantum walk-based algorithms to real-world scenarios. However, practical challenges, such as achieving sufficient spatial resolution and stability and scaling to larger system sizes, must still be addressed. Future work will need to explore these technical aspects to enable the successful experimental realization of quantum walk-based algorithms.

In summary, the use of potential functions in quantum walks presents a powerful framework for both algorithmic development and modeling noise in quantum systems. Further analytical and numerical studies, especially those focusing on more complex potential landscapes, will likely deepen our understanding of the role quantum walks can play in search and optimization problems.

\section*{Acknowledgements}
R.A.M.S. and F.L.M. thank the financial support from the Latvian Quantum Initiative under the European Union Recovery and Resilience Facility project number 2.3.1.1.i.0/1/22/I/CFLA/001. F.L.M. thanks the financial support from CNPq/Brazil grant number 407296/2021-2.

\section*{References}

\bibliographystyle{unsrt}
\bibliography{references}

\end{document}